\documentclass[aps,prl, amsmath, showpacs, superscriptaddress, twocolumn,sort&compress,floatfix, amssymb, noeprint]{revtex4-1}
\usepackage{graphicx,color}
\usepackage{mathptmx, textcomp, float}
\usepackage[latin1]{inputenc}
\usepackage{braket,amsfonts}
\usepackage{hyperref}
\usepackage{multibib}

\hypersetup{
    colorlinks = false,
    hidelinks = true,
}
\newcommand{\sech} { \mathrm{sech} }

\begin{document}

\title{Excitation modes of bright matter-wave solitons}
\author{Andrea Di Carli, Craig D. Colquhoun, Grant Henderson, Stuart~Flannigan, Gian-Luca Oppo,
Andrew J. Daley, Stefan Kuhr, Elmar Haller}
\affiliation{Department of Physics and SUPA, University of Strathclyde, Glasgow G4 0NG, United Kingdom}

\date{\today}

\begin{abstract}
We experimentally study the excitation modes of bright matter-wave solitons in a quasi-one-dimensional geometry. The solitons are created by quenching the interactions of a Bose-Einstein condensate of cesium atoms from repulsive to attractive in combination with a rapid reduction of the longitudinal confinement. A deliberate mismatch of quench parameters allows for the excitation of breathing modes of the emerging soliton and for the determination of its breathing frequency as a function of atom number and confinement. In addition, we observe signatures of higher-order solitons and the splitting of the wave packet after the quench. Our experimental results are compared to analytical predictions and to numerical simulations of the one-dimensional Gross-Pitaevskii equation.
\end{abstract}

\maketitle

The dispersionless propagation of solitary waves is one of the most striking features of nonlinear dynamics, with multiple applications in hydrodynamics, nonlinear optics and broadband long-distance communications \cite{Dauxois2006}. In fiber optics, one-dimensional (1D) ``bright'' solitons, i.e.\,solitons presenting a local electric field maximum with one-dimensional propagation, have been observed \cite{Mollenauer1980}. They exhibit a dispersionless flow and excitation modes such as breathing or higher-order modes \cite{Satsuma1974,Mollenauer1980,Stolen1983}. Matter waves can also display solitary dispersion properties. Typically, bright matter-wave solitons are created in quasi-1D systems by quenching the particle interaction in a Bose-Einstein condensate (BEC) from repulsive to attractive \cite{Khaykovich2002}. Recent experiments demonstrated the collapse \cite{Donley2001doc}, collisions \cite{Nguyen2014}, reflection from a barrier \cite{Marchant2013}, and the formation of trains \cite{Strecker2002, Nguyen2017, Gosar2019} of bright solitons.

In this letter, we experimentally study the excitation modes of a single bright matter-wave soliton. In previous studies, other dynamical properties have been observed, such as center-of-mass oscillations of solitons in an external trap \cite{Nguyen2014}, excitations following the collapse of attractive BECs \cite{Donley2001doc,Cornish2006}, and quadrupole oscillations of attractive BECs in three dimensions (3D) \cite{Everitt2018}. Here, we probe the fundamental breathing mode of a single soliton by measuring its oscillation frequency and the time evolution of its density profile. In addition, we observe signatures of higher-order matter-wave solitons, which can be interpreted as stable excitations with periodic oscillations of the density profile and phase, or as a bound state of overlapping modes \cite{Satsuma1974,Carr2002}.

The shape-preserving evolution of a matter-wave soliton is due to a balancing of dispersive and attractive terms in the underlying 3D Gross-Pitaevskii equation (GPE) \cite{Dalfovo1999c}. For quasi-1D systems with tight radial confinement, we can approximate the matter wave in the 3D-GPE by the product of a Gaussian wave function for the radial direction and a function $f(z)$ for the longitudinal direction (see \cite{SuppMat}). Depending on the ansatz for the Gaussian with either constant or varying radial sizes, $f(z)$ satisfies either the 1D-GPE or the non-polynomial Schr\"odinger equation (NPSE) \cite{Salasnich2002}. We reference to the analytical solutions of the 1D-GPE in the manuscript, but use both equations in our numerical simulations \cite{SuppMat}.

\begin{figure}[t]
\centering
  \includegraphics[width=0.49\textwidth]{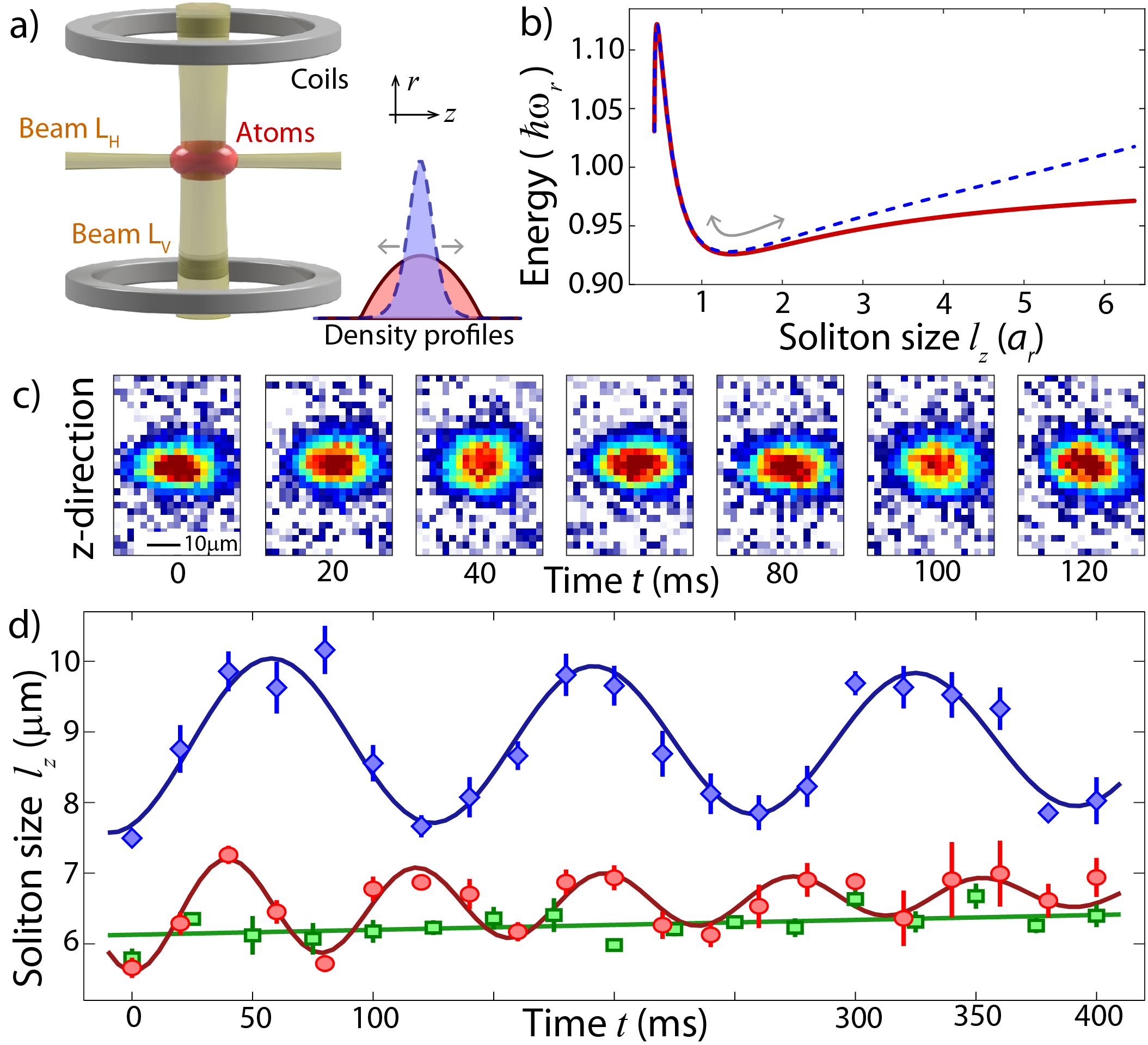}
  \vspace{-3ex}
 \caption{Experimental setup and oscillation measurements. (a) Sketch of the experimental setup. Inset: density profiles for a BEC (solid red line) and for a soliton (dashed blue line). (b) Total energy of a soliton, $a=-5.2\,a_0$, $\omega_r=2\pi\times 95$\,Hz, $N=2000$, with an external trap, $\omega_z=2\pi\times 5$\,Hz (dashed blue line), and without external trap, $\omega_z=0$\,Hz (solid red line). (c) Absorption images after a free expansion time of 16\,ms (from data set with circles in d), integrated density profile for $t=60\,$ms (blue line) and fit (dashed red line). (d) Oscillations of a quantum gas after the steps of a quench procedure. Blue diamonds: quench of only $\omega_z$ for a BEC (Q1), Red circles: additional interaction quench to create soliton (Q3), Green squares: optimized quench parameters to minimize breathing of the soliton (Q2). Uncertainty intervals indicate $\pm1$ standard error. \vspace{-3ex}\label{Fig:setup}}
\end{figure}

For the 1D-GPE, an ansatz for the normalized longitudinal wave function $f(z)$ is of the form
\begin{align} \label{eq:soliton}
    f(z) = \frac{1}{\sqrt{2 l_z}}\sech\left( \frac{z}{l_z} \right),
\end{align}
with a single parameter, $l_z$, that determines both the longitudinal size and the amplitude of the soliton. Solitons form with a value of $l_z$ that minimizes the total energy and that provides a compromise between the kinetic and the interaction energies. This is illustrated in Fig.\,\ref{Fig:setup}b, which shows the energy of the wave packet for varying sizes $l_z$ \cite{Parker2007b}. The kinetic energy provides a potential barrier for small $l_z$  that prevents the collapse of the soliton, while its spreading is inhibited by the interaction energy, which increases for large $l_z$.

Even without an external longitudinal potential, the soliton is stable against small perturbations of $l_z$. In a way, a bright matter-wave soliton creates its own trapping potential, which defines its size and excitation modes. Variational methods provide accurate predictions of its size at the energy minimum which can be calculated analytically \cite{Carr2002,Longenecker2018} or numerically \cite{Parker2007b}. For the fundamental solution (order $n=1$) of the 1D-GPE with an atom number $N$, s-wave scattering length $a$ and radial trapping frequency $\omega_r$, the size $l_z$ corresponds to the healing length at the peak density of the soliton, i.e. $l_z^{(n=1)} = a_r^2/(N|a|)$ \cite{Carr2002,Parker2007b}. Here, $a_r=\sqrt{\hbar/(m\omega_r)}$ is the radial harmonic oscillator length. Small deviations of $l_z$ close to the energy minimum lead to oscillations of the soliton size. We use those oscillations resulting from an initial mismatch of $l_z$ to experimentally measure the self-trapping frequency of the soliton potential.

Our experimental starting point is a Bose-Einstein condensate of $500-2000$ cesium (Cs) atoms in the state $\ket{F=3,m_F=3}$ at scattering length of $a=+7\,a_0$, where $a_0$ is Bohr's radius. The BEC is levitated by a magnetic field gradient, and it is confined by an optical dipole trap formed by the horizontal and vertical laser beams $L_H$ and $L_V$ (Fig.\,\ref{Fig:setup}a).  An additional magnetic offset field allows us to tune the scattering length by means of a broad magnetic Feshbach resonance \cite{Berninger2013a}. Details about our experimental setup, the levitation scheme and the removal of atoms can be found in references \cite{DiCarli2019,SuppMat}.

Our matter-wave solitons are confined to a quasi-1D geometry with almost free propagation along the horizontal direction and strong radial confinement of $\omega_r=2\pi\times 95$\,Hz provided by laser beam $L_H$. They are generated with a quench of the scattering length towards attractive interaction ($a_i\rightarrow a_f$), and by a reduction of the longitudinal trap frequency ($\omega_{z,i}\rightarrow \omega_{z,f}$). When changing $a$ and $\omega_z$ independently, the quenches excite inward and outward motions, respectively. Usually, it is desirable to minimize the excitations of the soliton by matching the initial Thomas-Fermi density profile of the BEC closely to the density profile of the soliton (inset Fig.\,\ref{Fig:setup}a). However, we deliberately mismatch the quench parameters to create breathing oscillations of the soliton in order to study its self-trapping potential. Quenches with different parameters are labelled by the symbols Q1-Q7 (see \cite{SuppMat}). Following an evolution time $t$ in quasi-1D and after a short period of 16\,ms of expansion in free space, we take absorption images to determine the density profile of the atoms (Fig.\,\ref{Fig:setup}c). The cloud size, $l_z(t)$, is determined by fitting the function $A(\sech(z/B))^2$ to the integrated 1D-density profiles with fit parameters $A$ and $B$ \cite{SuppMat}.

The response of the atomic cloud to the different steps of the quench protocol is presented in Fig.\,\ref{Fig:setup}d. We first quench only the longitudinal confinement by 25\% to $\omega_{z,f} = 2\pi\times 4.3(2)$\,Hz (quench Q1 in \cite{SuppMat}) while keeping the repulsive interaction strength constant (Fig.\,\ref{Fig:setup}d, diamonds). The BEC starts an outwards motion with an oscillation frequency of $2\pi\times  7.5(1)\,\mathrm{Hz}\approx \sqrt{3}\omega_{z,f}$ as expected for a BEC in the Thomas-Fermi regime \cite{Menotti2002a,Haller2009b}. In a second step, we additionally quench the interaction strength $a_f$ to $-5.4\,a_0$ and increase $\omega_{z,i}$ to match the initial size of the BEC to the expected size of the soliton, (Q2, Fig.\,\ref{Fig:setup}d, squares). As a result, we observe almost dispersionless solitons with a linear increase of $l_z$ of $0.7(3)\,\mu$m/s (Fig.\,\ref{Fig:setup}d, green line). Finally, we deliberately mismatch the initial size of the BEC by reducing $\omega_{z,i}$ (Q3), and generate small amplitude oscillations of the soliton with a frequency $\omega_{sol}$ of $2\pi\times 12.8(4)\,$Hz (Fig.\,\ref{Fig:setup}d, circles). This breathing frequency of the soliton is significantly larger than any breathing frequency of a BEC or of non-interacting atoms, $2\omega_{z,f}= 2\pi\times 8.6(3)\,$Hz. We observe no discernible oscillation in the radial direction after the quenches.

\begin{figure}[t]
\centering
  \includegraphics[width=0.48\textwidth]{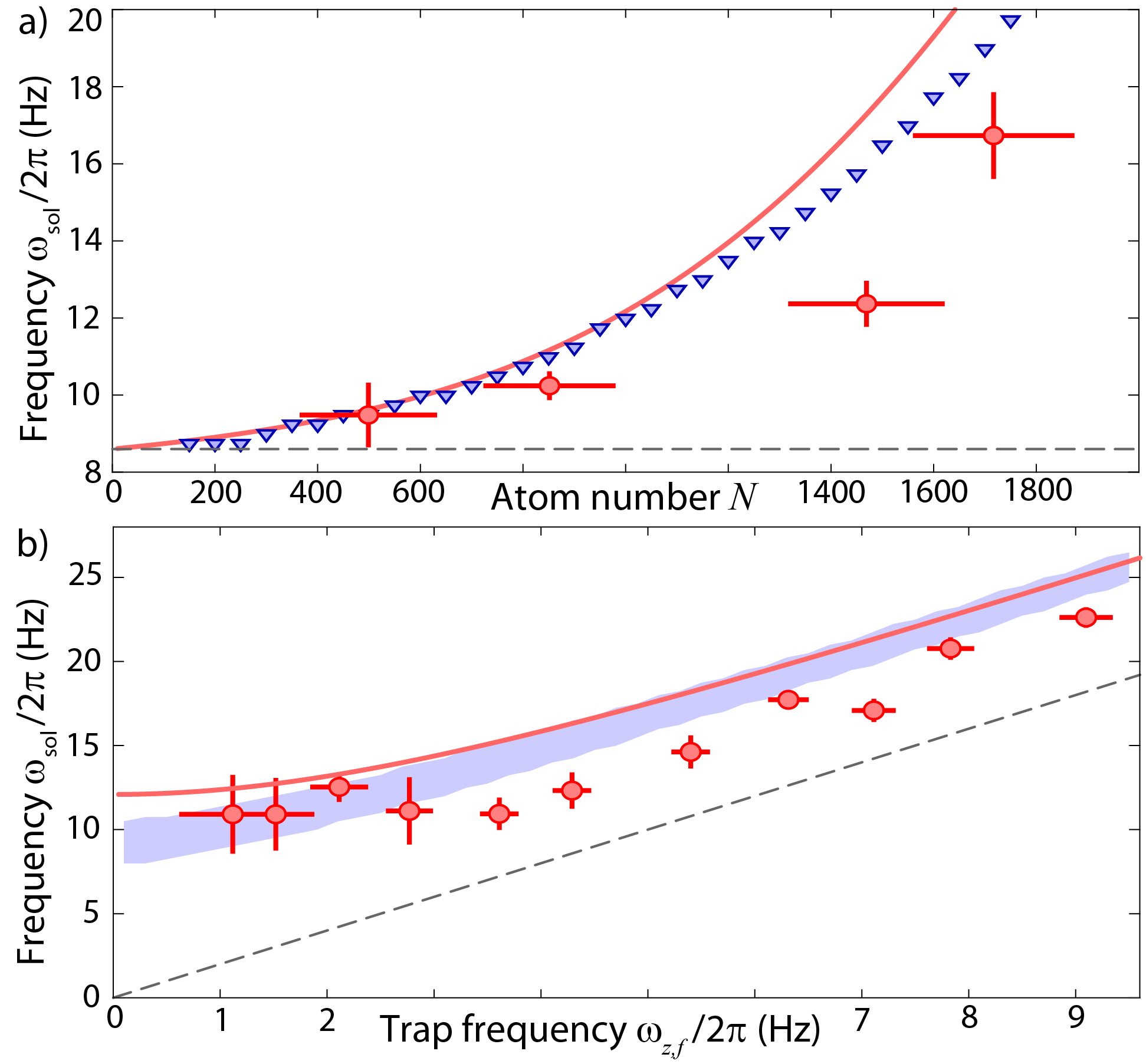}
  \vspace{-3ex}
  \caption{Breathing frequency $\omega_{sol}$ of the soliton. (a) Atom number dependence (Q4). Red circles:  experimental data, the uncertainty bars for the atom number indicate the standard deviation of $N$ over the first 100\,ms of each frequency measurement. Blue triangles: simulation of the 1D-GPE \cite{SuppMat}. Red line: analytical approximation \cite{Carr2002,SuppMat}. Dashed gray line: oscillation frequency of a non-interacting gas, $2\omega_{z,f}$. (b) Dependence of $\omega_{sol}$ on the trap frequency (Q5). Red circles: experimental data points for $N\approx1450$. Blue area: simulation of the 1D-GPE for $N=1300$ to $1500$. Red line: analytical approximation. Dashed gray line: $2\omega_{z,f}$. \vspace{-3ex}\label{Fig:breathingFreq}}
\end{figure}

In a second experiment, we demonstrate  that the breathing frequency, $\omega_{sol}$, depends on the interaction term $N a$ in the 1D-GPE, a property typical of the nonlinear character of the soliton. We choose to change $N$, since the initial removal process is independent of the interaction quench, and we can study $\omega_{sol}$ without changing the quench protocol (Q4, Fig.\,\ref{Fig:breathingFreq}a, circles). The measured values of $\omega_{sol}$ decrease for lower $N$, and they approach the breathing frequency $2\omega_{z,f}$ for non-interacting atoms in a harmonic trap (Fig.\,\ref{Fig:breathingFreq}a, dashed line).

We compare our experimental data points to two theoretical models. In a numerical simulation of the 1D-GPE, we use the ansatz in Eq.\,\ref{eq:soliton} to set the starting conditions, and we determine the breathing frequency from a spectral analysis of the time evolution of the wave function \cite{SuppMat} (Fig.\,\ref{Fig:breathingFreq}a, triangles). In addition, we use an analytical approximation for the breathing frequency (red line) calculated with a Lagrangian variational analysis at the energy minimum of the 3D-GPE \cite{Carr2002,SuppMat}). We find that both models agree well with the trend of the measurements of $\omega_{sol}$, although our experimental data points are systematically lower for large $N$ than our theoretical predictions. We speculate that this is due to non-harmonic contributions to the energy of the soliton on the breathing oscillations for finite oscillation amplitudes (Fig.\,\ref{Fig:setup}b).

To determine the influence of the trapping potential, we measure the variation of $\omega_{sol}$ as we reduce the longitudinal trapping frequency $\omega_{z,f}$ (Q5). Two regimes of $\omega_{sol}$ can be identified in Fig.\,\ref{Fig:breathingFreq}b for varying the values of $\omega_{z,f}$. For large values of $\omega_{z,f}$, the trap dominates the breathing of the soliton and $\omega_{sol}$ approaches twice the trap frequency $2\omega_{z,f}$. For small values of $\omega_{z,f}$, interactions dominate the breathing of the soliton and $\omega_{sol}$ reaches a constant value. This offset of the breathing frequency is a result of the ``self-trapping'' potential of a free soliton.

Again, we compare the experimental results with our theoretical model (Fig.\,\ref{Fig:breathingFreq}b, red line) and the numerical simulations of the 1D-GPE. The blue band in Fig.\,\ref{Fig:breathingFreq}b indicates the simulated frequencies for $N=1300$ to $N=1500$. The simulation predicts a lower breathing frequency for the free soliton than the analytical approximation, but all curves are within the uncertainly range of the experimental data.

External trapping potentials can in principle alter the soliton dynamics \cite{Serkin2015,Serkin2019a,Nguyen2014}, causing, e.g., modulations of the soliton's tails due to residual non-autonomous terms of the 1D-GPE in a harmonic potential \cite{Tenorio2005}. For the following experiments, however, we employ trap frequencies that are significantly smaller than the observed oscillation frequencies of the soliton ($2\omega_z<\omega_{sol}$) to decouple the influence of the trapping potential. In summary, for small-amplitude oscillations we find good agreement of $\omega_{sol}$ between our experimental results and analytical and numerical predictions based on the 1D-GPE (and NPSE \cite{SuppMat}).

Breathing oscillations of $l_z$ close to the equilibrium size are not the only possible excitation modes of solitons. The existence of higher-order solitons has been predicted in the nonlinear Schr\"odinger equation (NSE) \cite{Satsuma1974}, and has been observed for optical solitons in silica-glass fibers \cite{Mollenauer1980,Stolen1983}. A soliton of order $n$ can be interpreted as a bound state of $n$ strongly overlapping solitons \cite{Carr2002}. By exploiting the equivalence of the NSE and 1D-GPE, similar effects were later proposed for bright matter-wave solitons \cite{Carr2002,Golde2018a}, where it was suggested that n$^{th}$-order solitons can be generated by a rapid increase of the attractive interaction strength by a factor $n^2$. Similarly, our simulations of the 1D-GPE show that higher-order solitons can be created for an increased initial size of the wave packet. An $n^{th}$-order soliton forms for a sech-shaped wave function with an initial size $l_z^{(n)}$ that is the $n^2$ multiple of the healing length $l_z^{(1)}$, i.e. $l_z^{(n)} = n^2 \; l_z^{(1)}$ \cite{SuppMat}.

\begin{figure}[t]
\centering
  \includegraphics[width=0.48\textwidth]{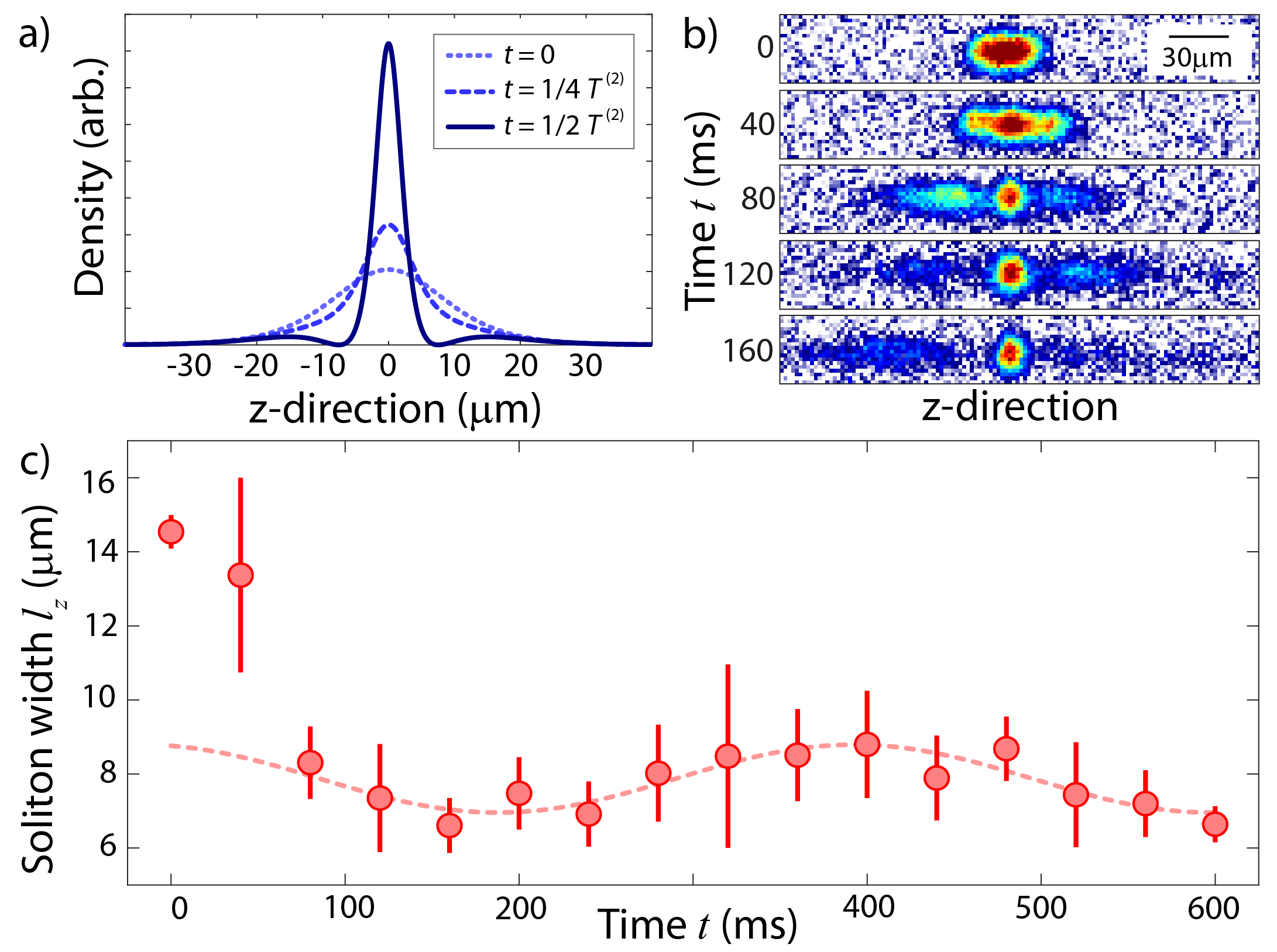}
  \vspace{-3ex}
  \caption{Time evolution after a strong quench of interactions and trap frequency (Q6). (a) Absorption images at time $t$ after the quench and after 11\,ms of free expansion. (b) 1D-GPE simulation of the density profiles for a second-order soliton with 1100 atoms, $a_f=-5.3\,a_0$, and with an oscillation period $T^{(2)}$ of 432\,ms. (c)  Time evolution of the measured width $l_z$ of the central wave packet (red circles), sinusoidal fit with period 420(30)\,ms (dashed red line). The uncertainty intervals indicate $\pm1$ standard deviation. \label{Fig:SingleQuench}
  \vspace{-3ex} }
\end{figure}

Within the 1D-GPE theory, both creation methods result in the periodic development of multi-peaked structures for higher-order solitons \cite{Satsuma1974, Gamayun2016}, e.g. they create a sharp central peak with side wings for a second-order soliton (Fig.\,\ref{Fig:SingleQuench}a) and a double peak for a third-order soliton \cite{SuppMat}. Sizes and interaction quenches that do not fulfil the previous conditions, lead to a ``shedding'' of the atomic density in the $z-$direction. The wave packet oscillates and loses particles until its size and shape match the next (lower $n$) higher-order soliton \cite{Satsuma1974}. For a second-order soliton, the predicted oscillation period $ T^{(2)}$ is \cite{Carr2002}
\begin{align} \label{eq:Period2ndOrder}
  T^{(2)} = \frac{8\pi}{\hbar} m \left(l_z^{(1)}\right)^2.
\end{align}

Recently, excitation modes of higher-order have also been used as a testbed for various theoretical models beyond GP-theory. The fragmentation of solitons with an increased initial width was predicted within the multiconfigurational time-dependent Hartree method for bosons \cite{Streltsov2008} and critically discussed \cite{Cosme2016}, and the influence of quantum effects on the dissociation process was investigated \cite{Weiss2016a,Yurovsky2017,Ng2019}.

Here, we apply two different quench protocols to study the evolution of strongly excited solitons. Depending on the initial size and the quench parameters, we observe shedding and fragmentation of the wave packet, and we measure oscillation frequencies that indicate the creation of higher-order solitons. To demonstrate the effect of a strong quench of an elongated BEC, we increase $a_i$ and reduce $\omega_{z,i}$ before ramping $a$ and $\omega_{z}$ to $-5.3\,a_0$ and $2\pi\times 0.0(6)\,$Hz in $13\,$ms (Q6).  Our quench induces an initial spreading of the wave packet, followed by a strong shedding of atoms, and finally, in the formation of a soliton that contains approximately 1/3 of the initial atom number (Fig.\,\ref{Fig:SingleQuench}b). We determine the soliton width and find a slow oscillation of $l_z(t)$ with a frequency of $2\pi\times 2.4(2)$\,Hz (Fig.\,\ref{Fig:SingleQuench}c). This frequency is significantly smaller than the expected breathing frequency of first-order solitons, $2\pi\times 6.0$\,Hz, and it matches well to the expected frequency of $2\pi\times 2.3$\,Hz for second-order solitons in Eq.\,\ref{eq:Period2ndOrder}.

\begin{figure}[t]
\centering
  \includegraphics[width=0.48\textwidth]{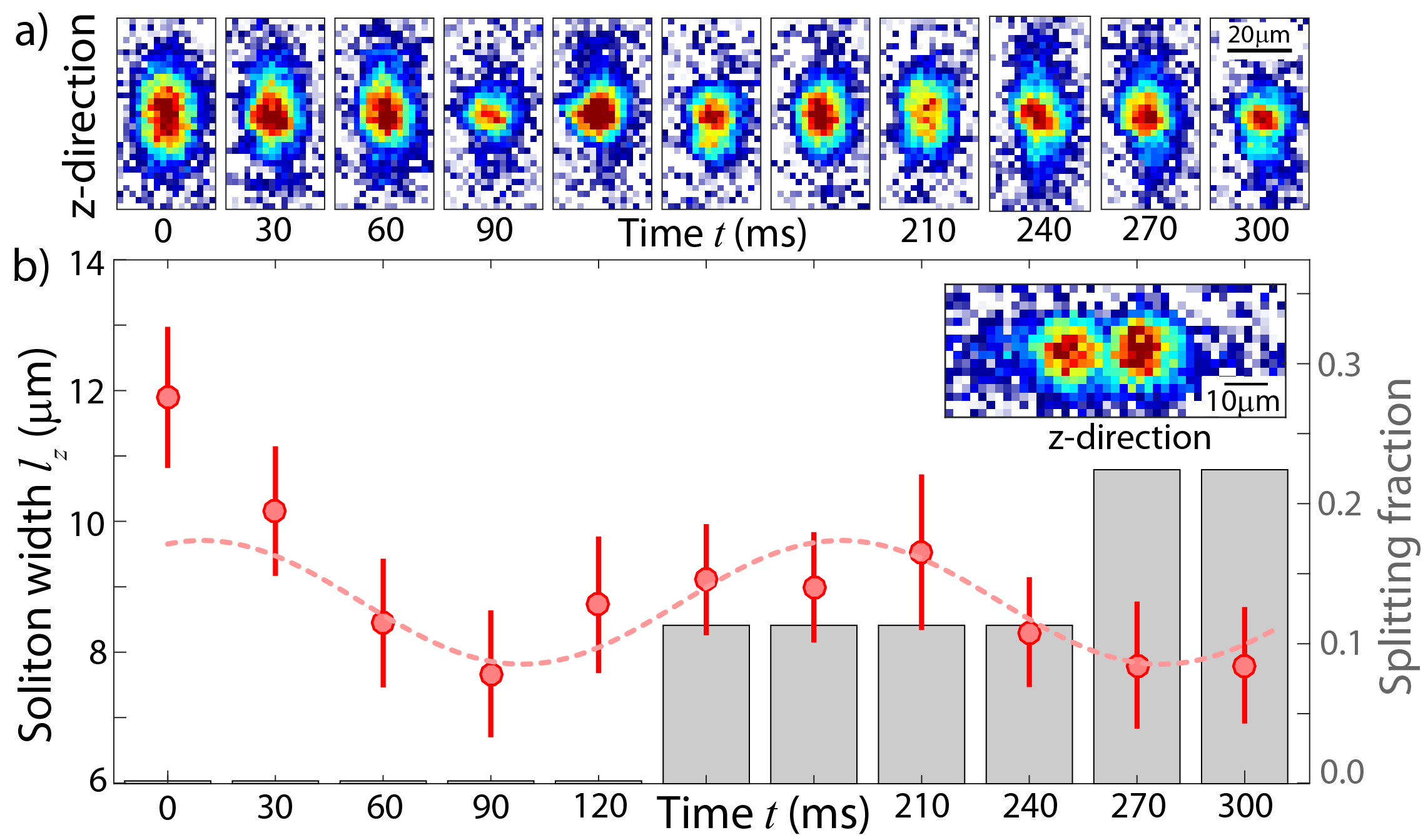}
  \vspace{-3ex}
    \caption{Second-order soliton and splitting after the double quench Q7. (a) Absorption images at time $t$ after the quench and after 7\,ms of free expansion. (b) Time evolution of the measured width $l_z$ of the central wave packet (red circles), sinusoidal fit with period 180(20)\,ms (dashed red line). The expected period from the 1D-GPE simulations is 192\,ms. The histogram counts the fraction of images showing a splitting of the wave function (9 repetitions per time step).  Inset: absorption image of a split matter wave for $t=210\,$ms. \label{Fig:DoubleQuench} \vspace{-3ex}}
\end{figure}

Observing shedding and oscillations agrees well with the predictions for higher-order solitons within the 1D-GPE \cite{Satsuma1974}, however, we find a strong dependence on details of the quench protocol and on the dynamical evolution during the quench. 
For a closer match to theoretical works \cite{Carr2002}, we implement a double-quench protocol, with a first quench to generate a soliton with weak attractive interaction, $a_{f}=-0.8\,a_0$, $\omega_f=2\pi\times 1.4(2)\,$Hz, and, after a settling time of $25\,$ms, a second quench of only the interaction strength, $a_{f}=-4.6\,a_0$ (Q7). Starting with approximately 2200 atoms, we observe no shedding but a small loss of 300 atoms during the first 60\,ms. The density distributions (Fig.\,\ref{Fig:DoubleQuench}a) resemble the expected profiles of a second-order soliton (Fig.\,\ref{Fig:SingleQuench}a), and the width of the wave packet oscillates with a frequency of $2\pi\times 5.6(6)$\,Hz (Fig.\,\ref{Fig:DoubleQuench}b), which matches the expected frequency of $2\pi\times 5.2$\,Hz for second-order solitons ($2\pi\times 13.2$\,Hz for the first-order).

For both measurements (Fig.\,\ref{Fig:SingleQuench}c,\,\ref{Fig:DoubleQuench}b), a small percentage of absorption images show the splitting of the soliton into two fragments (inset Fig.\,\ref{Fig:DoubleQuench}b). Due to the destructive nature of our absorption images it is difficult to conclude on the evolution and on the cause of the splitting process. A double-peak structure in the density profile can indicate the generation of a third-order soliton, fragmentation due to quantum effects, or simply an insufficient technical control of our quench parameters. For our setup, the control of horizontal magnetic field gradients to avoid longitudinal accelerations is especially challenging \cite{DiCarli2019}. The percentage of images that show a splitting of the wave packet increases for longer evolution times, and we indicate their fraction in Fig.\,\ref{Fig:DoubleQuench}b with a histogram.

In conclusion, we experimentally studied the creation and the excitation of breathing modes of bright matter-waves solitons in a quasi-one-dimensional geometry after a quench of interaction and longitudinal confinement. We measured the ``self-trapping'' frequency $\omega_{sol}$ for first-order solitions and its dependence on $N$ and $\omega_z$. For stronger excitations and for a double-quench protocol, we observed signatures of second-order solitons and the shedding and splitting of the wave function. Further measurements of the splitting process and the damping of the oscillations due to shedding are necessary to distinguish technical fluctuations from higher-order solitons and fragmentation due to quantum effects  \cite{Weiss2016a,Yurovsky2017,Ng2019}.

We thank L. D. Carr for helpful initial discussions during his visit. We acknowledge support by the EU through ``QuProCS'' (GA 641277) and the ETN ``ColOpt" (GA 721465), and by the EPSRC Programme Grant ``DesOEQ" (EP/P009565/1).  AdC acknowledges financial support by EPSRC and SFC via the IMPP.

\bibliography{SolArt,Manual,SuppMaterial}

\onecolumngrid\newpage\twocolumngrid

\section*{Supplemental Material}

\section{Experimental methods}

\subsection{Controlling the atom number in the BEC}
The solitons are confined to a quasi-1D geometry with almost free propagation along the horizontal direction and strong radial confinement of $\omega_r=2\pi\times 95$\,Hz provided by laser beam $L_H$. In quasi-1D geometry, bright matter-wave solitons collapse for large densities and interactions \cite{Donley2001doc}, which for our typical experimental scattering length of approximately $-5\,a_0$ corresponds to a critical atom number of 2500 \cite{Carr2002}. As a result, we need to strongly reduce the atom number to avoid collapse, modulation instabilities \cite{Strecker2002} and three-body loss \cite{Everitt2017a} for a deterministic and reproducible creation of the soliton. We remove atoms with a small additional magnetic field gradient, which pushes the atoms over the edge of the optical dipole trap. Our precise control of magnetic field strengths allows us to reduce the atom number down to $200$ atoms, with a reproducibility of $\pm 100$ for 600 atoms and $\pm350$ for 4500 atoms, measured as the standard deviation of the atom number in 50 consecutive runs. A removal period of 4\,s and smooth ramps of the magnetic field strength are necessary to minimize excitations of the BEC. Following the removal procedure we measure residual fluctuations of the width of the BEC below 3.5\%.\\

\subsection{Quench parameters}
Several different quench protocols are employed for the measurements. The quenches are labeled by the symbols Q1-Q7 in the main article:

\begin{itemize}
    \item[Q1] We quench only the trap frequency from $\omega_{z,i}=2\pi\times 5.8(2)$\,Hz to $\omega_{z,f}=2\pi\times 4.3(2)$\,Hz with a linear ramp of the laser power of beam $L_V$ over $4\,$ms. Atom number $N\approx1800$, constant interaction strength $a_i=+7\,a_0$, $\omega_r = 2\pi\times $95\,Hz.
    \item[Q2] In addition to the quench Q1 of the trap frequency, we also quench the interaction strength from $a_i=+7\,a_0$ to $a_f=-5.4\,a_0$ in 4\,ms. We minimize oscillations of the width of the soliton by reducing the initial size of the BEC with $\omega_{z,i} = 2\pi\times 11.2(2)\,$Hz. Atom number $N\approx1800$, $\omega_{z,f}=2\pi\times 4.3(2)$\,Hz, $\omega_r = 2\pi\times $95\,Hz.
    \item[Q3] We mismatch the initial size of the BEC before the quench with $\omega_{z,i} = 2\pi\times 12.8(4)$\,Hz to generate small amplitude oscillations of the width of the soliton. Atom number $N\approx1700$, $\omega_{z,f}=2\pi\times 4.3(2)$\,Hz, $\omega_r = 2\pi\times $95\,Hz.
    \item[Q4] Quench to determine the atom-number dependence of $\omega_{sol}$. We vary the atom number $N$ from 500 to 1700 for the measurement. $\omega_{z,f}=2\pi\times 4.3(2)$\,Hz, $a_i = +7\,a_0$, $a_f = -5.4\,a_0$, ramp duration 4\,ms, $\omega_r = 2\pi\times $95\,Hz.
    \item[Q5] Quench to determine the dependence of $\omega_{sol}$ on the trap frequency $\omega_{z,f}$. We vary $\omega_{z,f}$ from approximately 1\,Hz to 9\,Hz.
    Smaller values of $\omega_{z,f}$ result in larger equilibrium sizes of the soliton, and we need to reduce the initial trap frequencies $\omega_{z,i}$ to keep the oscillation amplitudes comparable during the measurements.     The typical difference between $\omega_{z,i}$ and $\omega_{z,f}$ is approximately $3\,$Hz. $N\approx 1500$, $a_f=-5.4\,a_0$, $\omega_r = 2\pi\times $95\,Hz.
    \item[Q6] Strong quench starting from an elongated BEC to excite higher-order oscillations and shedding. The ratio between the calculated initial Thomas-Fermi radius of the BEC and the expected width $l_z$ of the soliton is 24. $a_i=56\,a_0$, $a_f=-5.3\,a_0$, $\omega_{z,i} = 2\pi\times 4.9(2)$\,Hz, $\omega_{z,f} = 2\pi\times 0.0(6)$\,Hz, initial atom number $N\approx 3000$ drops to 1100 after shedding of atoms, quench duration $13\,$ms,  $\omega_r = 2\pi\times $86\,Hz.
    \item[Q7] Double quench to create a stable soliton in step 1 and quench the scattering length by approximately a factor of 4 in step 2. Step 1: $\omega_{z,i} = 2\pi\times 4.9(2)$\,Hz, $\omega_{z,f} = 2\pi\times 1.4(2)$\,Hz, $a_i=29\,a_0$, $a_f=-0.8\,a_0$, $\omega_r = 2\pi\times $86\,Hz, quench duration 15\,ms, $N\approx2200$. Settling delay of 25\,ms between quenches. Step 2: reduce interaction strength in 2\,ms to $a_f=-4.6\,a_0$, no change of other parameters.
\end{itemize}

\subsection{Fit of density profiles}
\begin{figure}[h]
\includegraphics[width=0.45\textwidth]{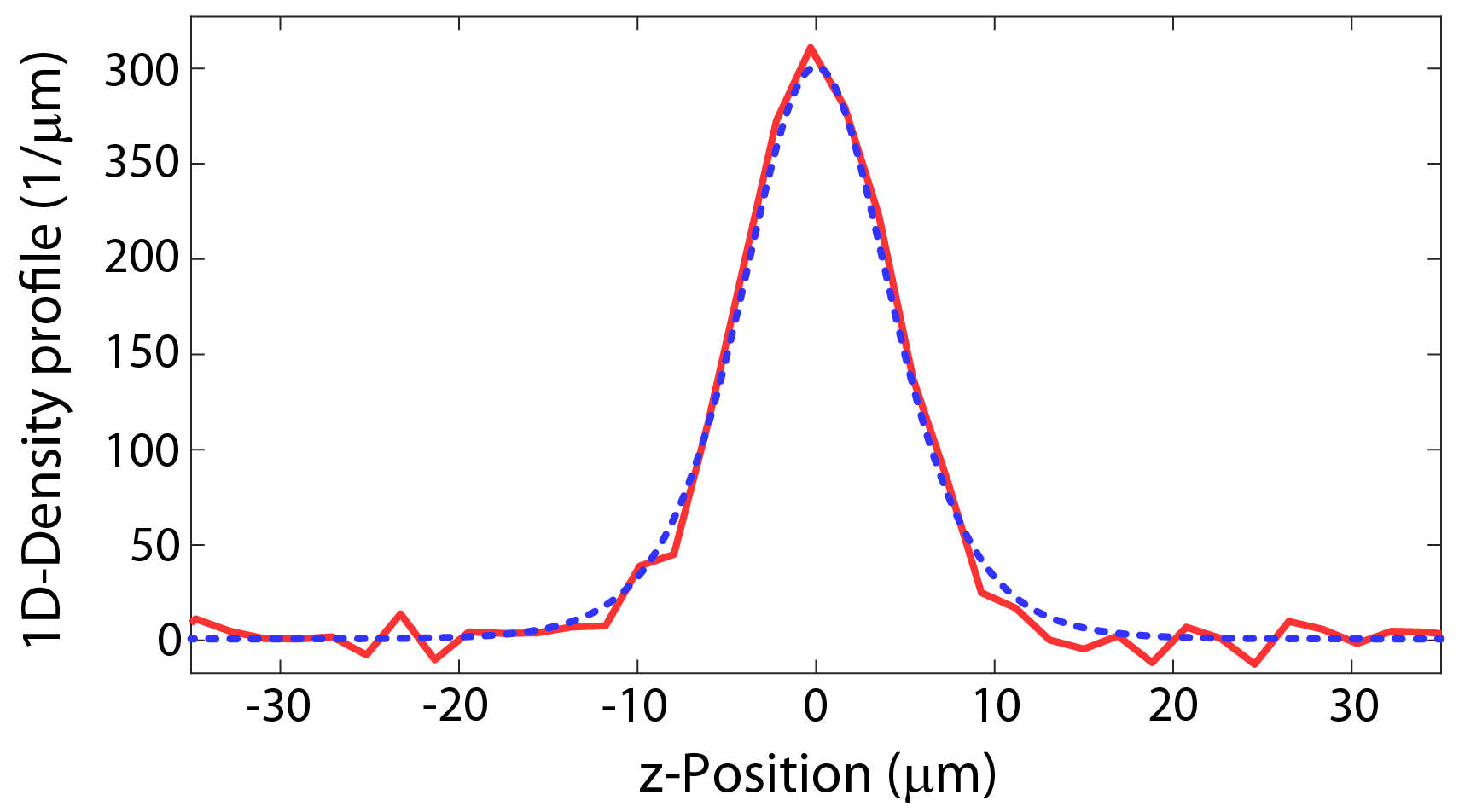}
\centering
\caption{1D-density profile of a soliton. Red line: integrated density profile of the absorption image for $t=60\,$ms in Fig.\,1c (main article). Dotted blue line: fitted profile according to Eq.\,1 in the main article.}\label{Fig:Fitting}
\end{figure}

We employ absorption imaging to measure the 2D-density profile of the soliton, and we integrate over one radial axis to determine the 1D-density profile (red line in Fig.\,\ref{Fig:Fitting}). The width $l_z$ of the soliton in Eq.\,1  of the main article, is determined by fitting the function $A(\sech(z/B))^2$, with fit-parameters $A$ and $B$, to the integrated 1D-density profiles (dotted blue line in Fig.\,\ref{Fig:Fitting}).

\section{Theoretical methods}

\subsection{The Model}

The time evolution of the collective wave function of $N$ atoms in an external potential with the 3D Gross-Pitaevski equation (GPE) for a time and space-dependent collective atomic wave-function, $\psi(\mathbf{r},t)$, is given by,
\begin{equation}\label{GPE_Diff}
i\hbar\frac{\partial}{\partial t}\psi(\mathbf{r},t)=\left[-\frac{\hbar^{2}}{2m}\nabla^{2}+V(\mathbf{r})+gN\left|\psi(\mathbf{r},t)\right|^{2}\right]\psi(\mathbf{r},t),
\end{equation}
where $g = 4\pi \hbar^2 a/ m$, $m$ is the atomic mass, and $a$ is the two-body s-wave scattering length.
This semi-classical field equation can be seen as a mean-field computation, and describes the dynamics of many weakly interacting particles at low temperatures when the condition $n |a|^3 \ll 1$ is satisfied \cite{PhysRevLett.88.210403}, where $n$ is the particle density. Our external potential $V(\mathbf{r})$ is given by a 3D (anisotropic) harmonic trap.

For tight radial trapping potentials, $\omega_r \gg \omega_z$, we can approximate the 3D wave function with a Gaussian solution in the radial directions and an arbitrary component, $f(z,t)$, in the longitudinal direction,
\begin{equation}
\psi(\mathbf{r},t)=f(z,t) \frac{1}{\sqrt{\pi}a_{r}\sigma(z,t)}\exp\left[-\frac{(x^{2}+y^{2})}{2a_{r}^{2}\sigma(z,t)^{2}}\right],
\end{equation}
where $a_r$ is the harmonic oscillator length in the radial direction and $\sigma(z,t)$ is a free parameter that dictates the width of the radial wavefunction. Substituting this ansatz into the 3D-GPE, and integrating over the radial directions, we arrive at the so called non-polynomial Sch\"odinger equation (NPSE) \cite{Salasnich2002}
\begin{equation}\label{PDE}
 \begin{split}
i\hbar \frac{\partial}{\partial t}f(z,t) &= \left[ -\frac{\hbar^{2}}{2m}\frac{\partial^{2}}{\partial z^{2}}+V(z) \right. \\
&+\frac{gN}{2\pi a_{r}^{2}\sigma(z,t)^{2}}\left|f(z,t)\right|^{2} \\
&\left. +\frac{\hbar\omega_{r}}{2}\left(\sigma(z,t)^{2}+\frac{1}{\sigma(z,t)^{2}}\right) \right] f(z,t),
 \end{split}
\end{equation}
where $\omega_r={\hbar/m a_r^2}$. The condition for $\sigma(z,t)$ that minimizes the action functional integrated along the trajectories in phase space is \cite{Salasnich2002},
\begin{equation}\label{sigE}
\sigma(z,t)^2 = \sqrt{1 + 2aN|f(z,t)|^2}.
\end{equation}

For $\sigma(z,t)=1$, we obtain the ground state of a harmonic oscillator in the radial directions, and we recover the usual 1D-GPE
\begin{equation}\label{GPE1D}
 \begin{split}
i\hbar \frac{\partial}{\partial t}f(z,t) &= \left[ -\frac{\hbar^{2}}{2m}\frac{\partial^{2}}{\partial z^{2}}+V(z) \right. \\
&  \left. +\frac{gN}{2\pi a_{r}^{2}}\left|f(z,t)\right|^{2}  + \hbar\omega_{r}  \right] f(z,t).
 \end{split}
\end{equation}

We have numerically integrated Eqs~\ref{PDE} and \ref{GPE1D} using the split-step Fourier transform method \cite{SSFF}, where we exploit the fact that the kinetic and potential terms in the Hamiltonian are diagonal in momentum and real space, respectively.

\section{Soliton Breathing Frequency}

\begin{figure}[h]
\includegraphics[width = 8cm]{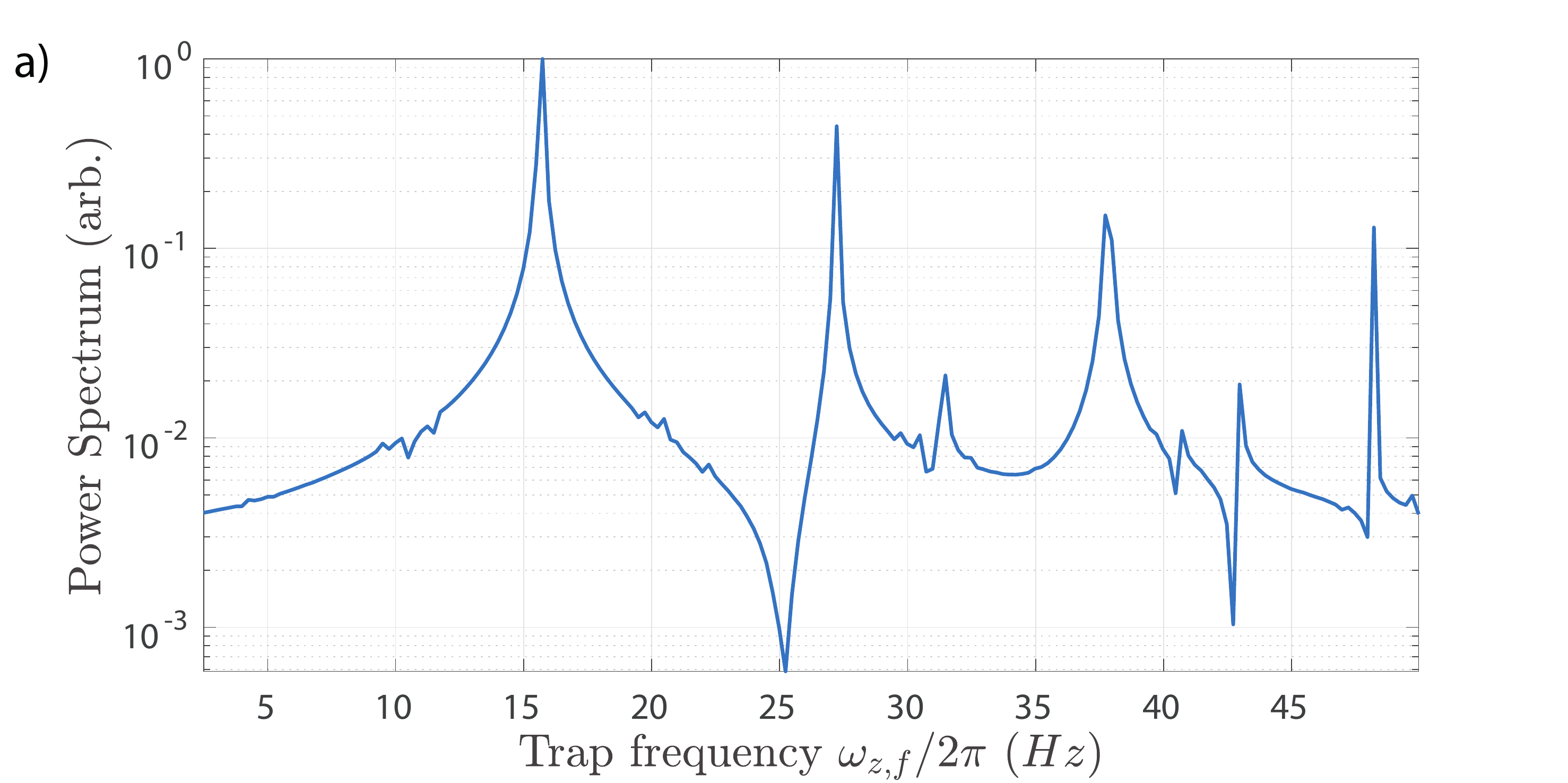}
\includegraphics[width = 8cm]{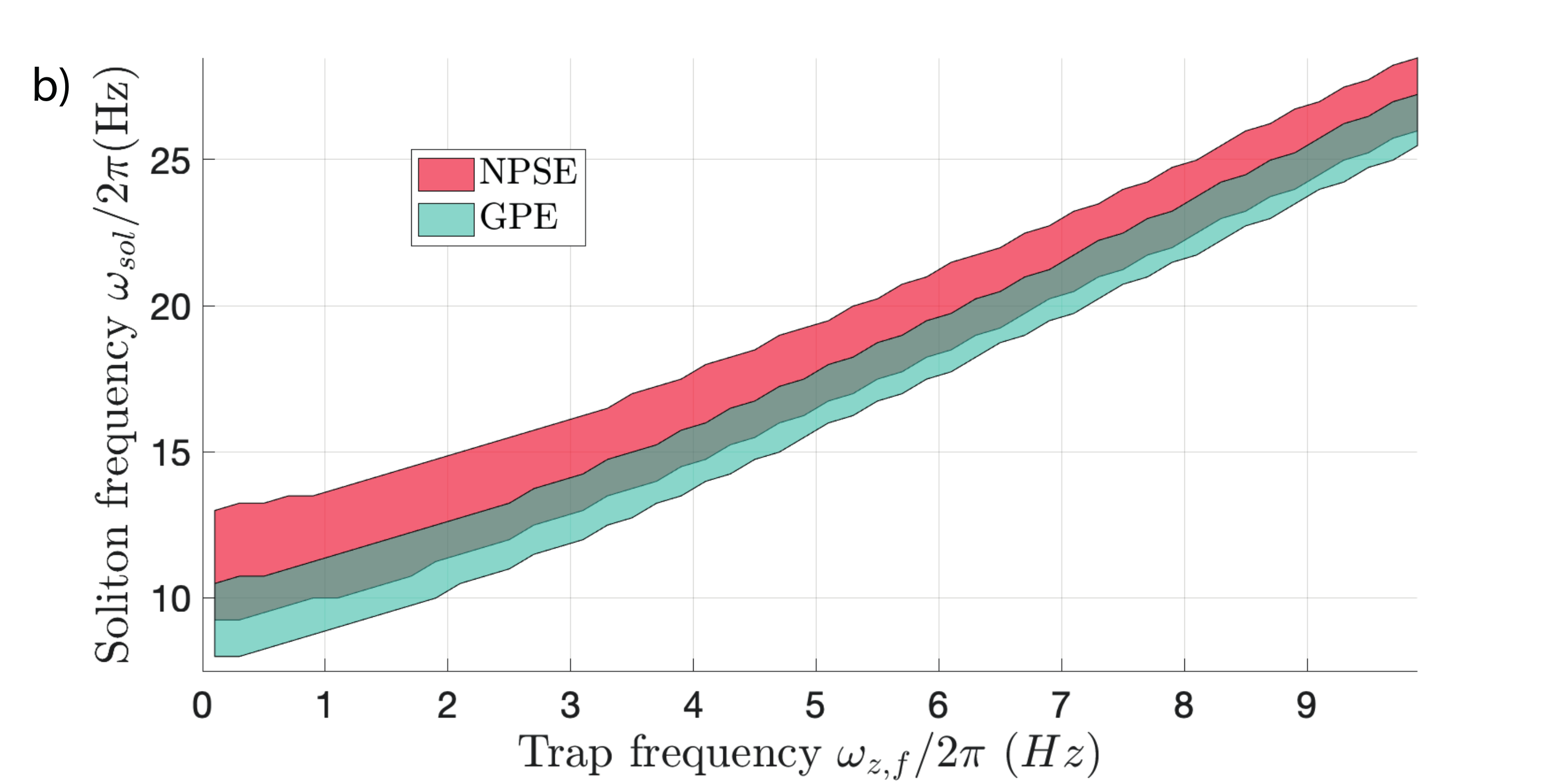}
\centering
\caption{Simulation results for the soliton breathing frequency, for comparison with Fig.\,2 in the main text. (a) Frequency spectrum calculated using the 1D-GPE with a longitudinal frequency of $\omega_z = 2\pi \times 5$\,Hz and atom number $N=1300$. (b) Breathing frequency (first peak in the spectrum as in a)) vs. trap frequency. $N \approx 1300 - 1500$ atoms, $a_f = -5.4$ $a_0$ for the NPSE (red) and the GPE (green). The simulations were evolved in time to $4000$\,ms.}
\label{sol_bre}
\end{figure}

In this section we explain how the numerical calculations of the soliton breathing frequencies shown in Fig.\,2 of the main text were carried out. We begin with the order 1 soliton solution,
 \begin{equation}\label{soliton}
f(z,0) = \frac{1}{\sqrt{2l_z}} \sech \left(\frac{z}{l_z}\right)
\end{equation}
where $l_z=a_r^2/(N|a_i|)$, and we have used $a_i=7\,a_0$. We then evolve this initial state either with the 1D-GPE or NPSE to a simulation time of $4000$\,ms and evaluate the frequency spectrum of the oscillation of the soliton's centre ($z=0$). In Fig.\,\ref{sol_bre} we present the frequency spectrum for the GPE and a longitudinal frequency of $\omega_z = 2\pi \times 5$\,Hz and atom number $N=1300$, which is characteristic of the behaviour for all other $\omega_z$ data points.  We observe several prominent frequency modes in the signal, but we select the lowest frequency peak to compare to the experimental measurements, because the resolution in the experiment is restricted to low frequency components.

Fig.\,\ref{sol_bre}b also shows the results of the simulation using both the 1D-GPE and the NPSE (compare with Fig.\,2 of the main text). We can see that for these atom numbers there are differences between the predictions of the 1D-GPE and NPSE. However these differences are small compared to the uncertainty in the experimental results.

\section{Higher order solitons}

\begin{figure}[t]
\centering
  \includegraphics[width=0.5\textwidth]{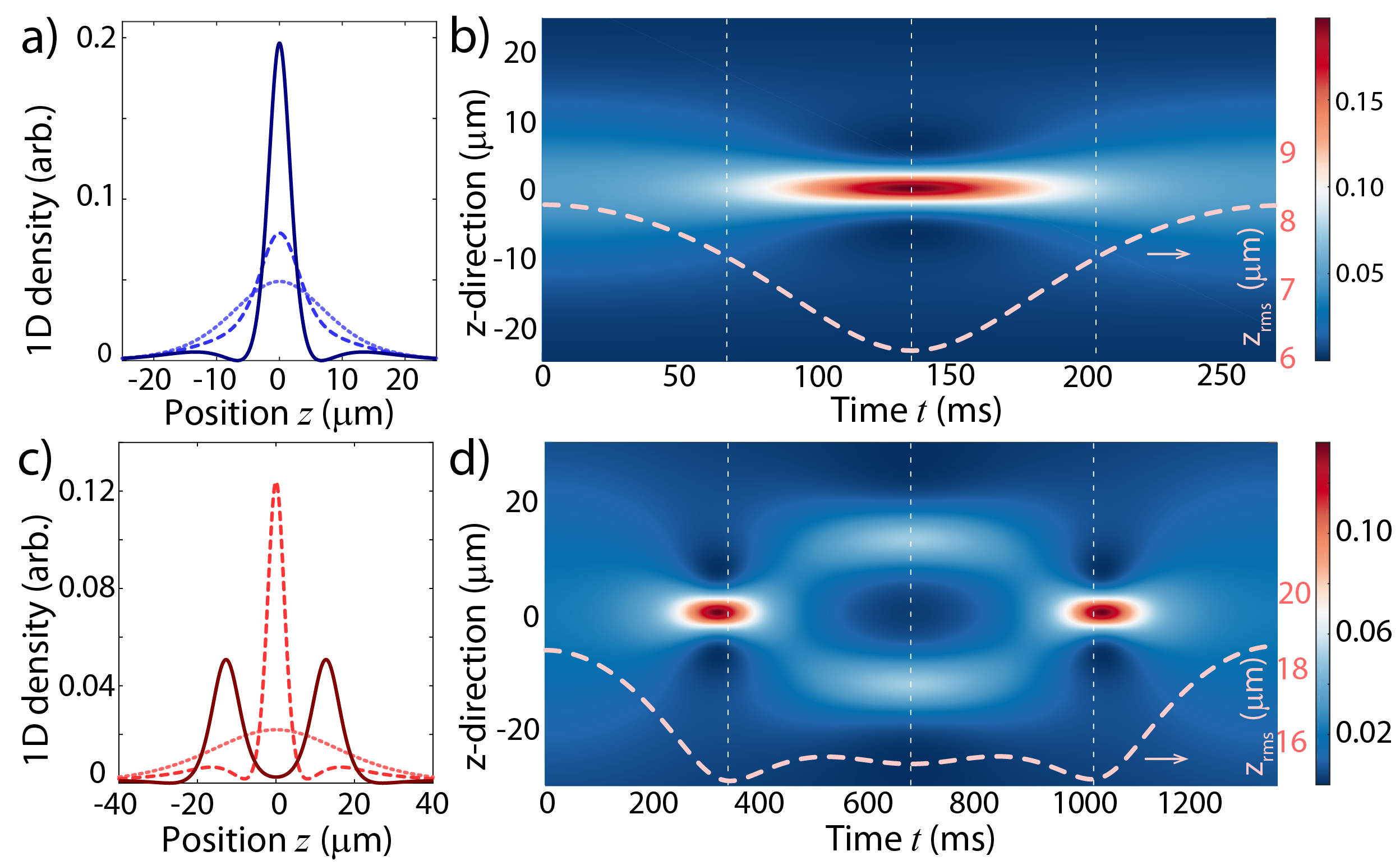}
  \vspace{-2ex}
\caption{Simulation of higher-order solitons in the 1D-GPE. Temporal snapshots (a) and temporal evolution (b) of the atomic density profile of an $n=2$ soliton for $N=1800$, $a=-3.7\,a_0$, $l_z^{(2)} = 10.2~\mu$m $=4 l_z^{(1)}$, and an oscillation period of $T_2 = 271\,$ms. Temporal snapshots (c) and temporal evolution (d) of the atomic density profile of an $n=3$ soliton for the same values of $N, a$, but with $l_z^{(3)} = 22.8~\mu$m$=9 l_z^{(1)}$, and with a period $T_3 = 1373\,$ms. The density profiles in (a) and (c) are plotted at $t=0$ (dotted lines), $t=1/4T$ (dashed lines), $t=1/2T$ (solid lines). The dashed lines in (b) and (d) display the temporal evolution of the size of the soliton wavepacket $z_{rms}$ (right scale). \label{Fig:HigherOrder}}
\end{figure}

Figure~\ref{Fig:HigherOrder} shows numerical simulations of the 1D-GPE for the time evolution of second- and third-order solitons with initial sizes $l_z^{(2)}$ and $l_z^{(3)}$. Large initial soliton sizes lead to the periodic formation of local maxima and minima of the density profile. Striking characteristics of the time evolution are the periodic development of a sharp central peak with side wings for the second-order soliton (Fig.\,\ref{Fig:HigherOrder}a,b), and the periodic formation of a broad double-peak structure for the third-order soliton (Fig.\,\ref{Fig:HigherOrder}c,d).

\begin{figure}[h]
\centering
\includegraphics[width = 8cm]{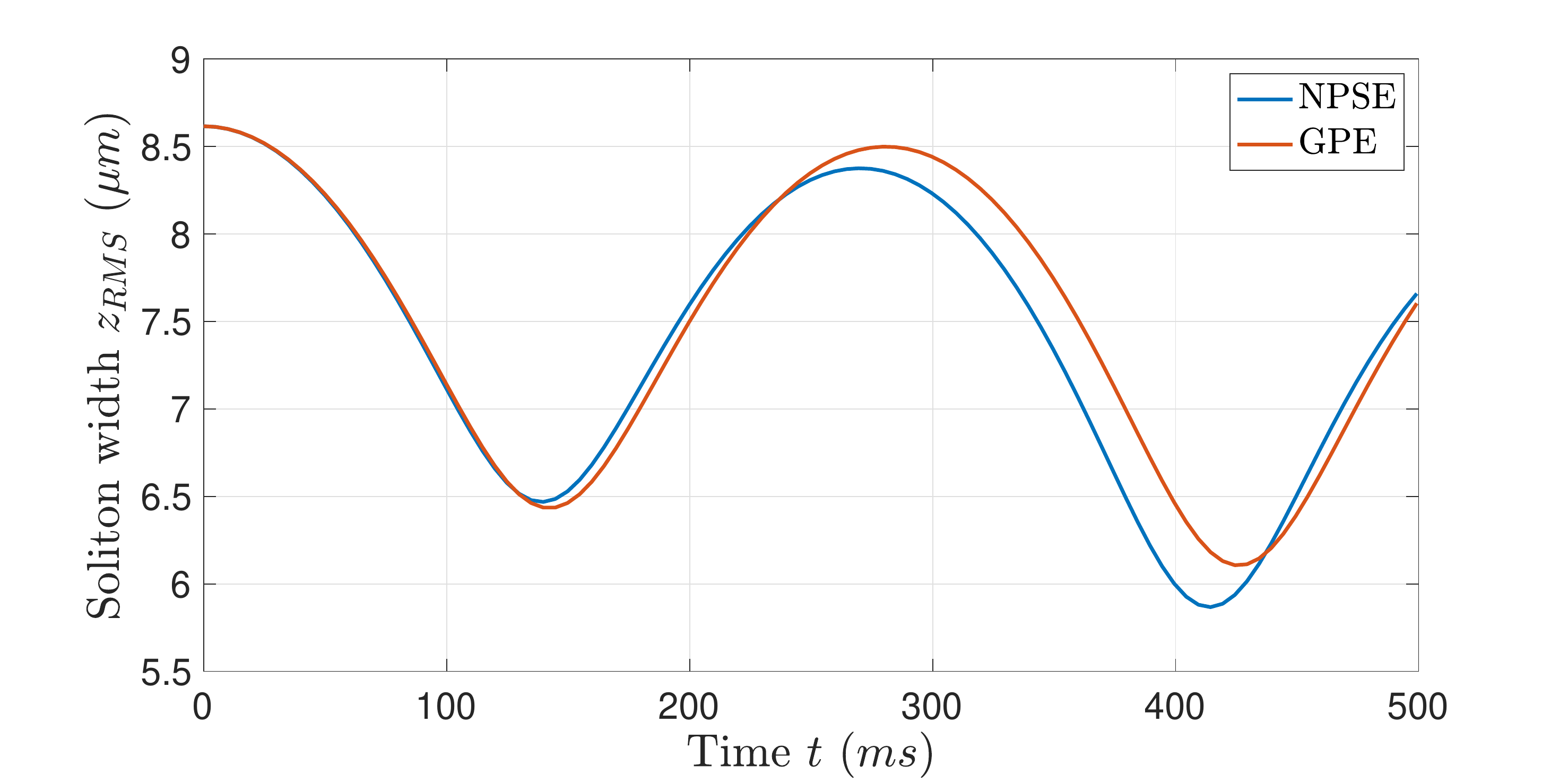}
\caption{Simulation results for the root mean square width of the soliton as it undergoes second order solitary behaviour, for the NPSE (blue) and the 1D-GPE (red). Here, $\omega_z = 0 $\,Hz, with an atom number $N = 1800$ and a scattering length $a = -3.7$ $a_0$.}
\label{fig:NPSEvsGPE2ndOrder}
\end{figure}

We also simulate the time evolution of solitons with the same start conditions using the NPSE and analyse the results using the root mean square width of the wave packet for a quantitative comparison (Fig.\,\ref{fig:NPSEvsGPE2ndOrder})
\begin{align}
    z_{rms}(t) = \left( \frac{1}{N}\int n(z,t) (z-\bar{z})^2 dz \right)^{1/2}.
\end{align}
Here, $\bar{z}$ is the mean position of the wave packet and $n(z,t)$ is the 1D-density. We observe small quantitative differences between the two equations but the overall behaviour is very similar.

\section{Variational approach for the BREATHING FREQUENCY}
In this section, we show how the longitudinal breathing frequency plotted in Fig.~2 of the main article can be determined from the variational ansatz for the soliton. For a cylindrical cigar-shaped potential the energy functional of Eq.\,\ref{GPE_Diff} is given by \cite{Carr2002,Parker2007b}
\begin{multline}\label{energyfunctional}
    E[\psi]=\int d^3\mathbf{r} \left[ \frac{\hbar^2}{2m}{|\mathbf{\nabla}\psi(\mathbf{r})|}^2+  \right. \\
          \left. \frac{1}{2}m(\omega^2_r r^2+\omega^2_z z^2){|\psi(\mathbf{r})|}^2+\frac{gN}{2} |\psi(\mathbf{r})|^4 \right]
\end{multline}
The energy of a soliton can be determined with a variational method using the following ansatz for the wave function
\begin{align}\label{Ansatz}
    \psi(r,z) = \frac{1}{\sqrt{2 l_z}}\sech\left( \frac{z}{l_z} \right)\,\cdot\,  \frac{1}{\sqrt{\pi l_r}} \exp\left(-\frac{r^2}{2l_r^2}\right),
\end{align}
where the transverse width $l_r$ and longitudinal width $l_z$ are the variational parameters \cite{Carr2002,Parker2007b}. Combining Eqs.\,\ref{energyfunctional} and \ref{Ansatz}, and rescaling the variables by the transverse frequency $\omega_r$, provides an equation for the normalized energy of the soliton \cite{Carr2002}
\begin{equation}\label{energyscaled}
\varepsilon_{GP}=\frac{1}{2\gamma^2_r}+\frac{\gamma^2_r}{2}+\frac{1}{6\gamma^2_z}+\frac{\pi^2}{24}\lambda^2\gamma^2_z+
\frac{\alpha}{3\gamma^2_r\gamma_z},
\end{equation}
with $\varepsilon_{GP}=E/\hbar\omega_r$, $\gamma_{r}=l_r/\sigma_r$, $\gamma_{z}=l_z/\sigma_r$, $\lambda=\omega_z/\omega_r$, $\alpha=Na/\sigma_r$,  and $\sigma_r=\sqrt{\hbar/m\omega_r}$. We can simplify Eq.\,\ref{energyscaled} for our system with weak interactions and strong transverse confinement by neglecting variations of the radial soliton size, i.e. $\gamma_r=1$. The energy minimum is found by calculating the zero-crossing of the first derivative of Eq.\,\ref{energyscaled} with respect to $\gamma_z$
\begin{equation}\label{equationminimum}
\frac{\pi^2}{4}\lambda^2\gamma^4_z+\sqrt{\zeta}\gamma_z-1=0,
\end{equation}
where $\alpha=-|\alpha|=-\sqrt{\zeta}$. Eq.\,\ref{equationminimum} has been solved for an expulsive potential with $\omega^2_z<0$ \cite{Carr2002}. Here, we provide the solution for a trapping potential with $\omega^2_z>0$. The longitudinal size of the soliton $\gamma_z^*$ at the energy minimum is
\begin{equation}\label{gamma}
\gamma^*_z=\frac{F}{\sqrt{\zeta}},
\end{equation}
with
\begin{equation}\label{ef}
F=-\sqrt{\frac{G}{2}}+\frac{1}{2}\sqrt{-2G+\frac{4\sqrt{2}}{\pi^2\sqrt{G}}\left(\frac{\zeta}{\lambda}\right)^2},
\end{equation}
where
\begin{equation}\label{gi}
G=\frac{\Delta}{\pi^{\frac{4}{3}}}\left(\frac{\zeta}{\lambda}\right)^{\frac{4}{3}}-\frac{4}{3\pi^{\frac{2}{3}}}\frac{1}{\Delta}
\left(\frac{\zeta}{\lambda}\right)^{\frac{2}{3}},
\end{equation}
with
\begin{equation}\label{delta}
\Delta=\sqrt[3]{1+\sqrt{1+\frac{64\pi^2}{27}\left(\frac{\lambda}{\zeta}\right)^2}}.
\end{equation}

In order to find the oscillation frequency $\omega_z$ of the soliton, the equations of motion for the variational parameters are determined with a Lagrangian variational analysis \cite{Carr2002}
\begin{equation}
\label{secondequationofmotion2}
    \left(\frac{\pi^2}{12}\right)\ddot{\gamma}_z=\frac{\gamma^{-3}_z}{3}-\frac{\pi^2}{12}\lambda^2\gamma_z+\frac{\alpha}{3}\gamma^{-2}_z,
\end{equation}
where the time derivative is calculated with respect to the normalised time $\tau=\omega_r t$. Again, we have assumed that the radial size of the soliton is constant, i.e. $\gamma_r=1$.

For small deviations of the soliton size from its equilibrium value, we can write the solution as $\gamma_z=\gamma^*_z+\delta\gamma_z$, where $\gamma^*_z$ is the minimum given by Eq.\,\ref{gamma} and $\delta\gamma_z$ is a small deviation. A linear expansion of Eq.\,\ref{secondequationofmotion2} leads to the expression for the longitudinal breathing frequency
\begin{align}\label{breathing}
    \omega_z=\omega_r\sqrt{\frac{12}{\pi^2}\left({\gamma^*_z}^{-4}+\frac{\pi^2}{12}\lambda^2+\frac{2\alpha}{3}{\gamma^{*}_z}^{-3}\right)}.
\end{align}
\vspace{0ex}

We compare our experimental measurements of the breathing frequency of the soliton to the predictions of Eq.\,\ref{breathing} in Fig.\,2 of the main article (red line).


\end{document}